\documentclass[aps,prl,twocolumn,showpacs,superscriptaddress]{revtex4-1}
\usepackage{amsmath}
\usepackage{epsfig}
\usepackage{color}
\usepackage{endnotes}
\usepackage{natbib}
\usepackage[colorlinks,breaklinks]{hyperref}
\usepackage{nicefrac}
\usepackage{bm}% bold math
\usepackage{dcolumn}
\usepackage{graphics}% Include figure files
\usepackage{amsmath}
\usepackage{amssymb}
\usepackage{color}
\usepackage{changes}
\newcommand{\bra}{\langle}
\newcommand{\ket}{\rangle}

\newcommand{\bs}[1]{\ensuremath{\boldsymbol{#1}}}

\newcommand{\be}{\begin{equation}}
\newcommand{\ee}{\end{equation}}
\newcommand{\bea}{\begin{align}}
\newcommand{\eea}{\end{align}}
\newcommand{\beqa}{\begin{eqnarray}}
\newcommand{\eeqa}{\end{eqnarray}}

\begin{document}

\title{The Generalized Nuclear Contact and its Application to
the Photoabsorption Cross-Section}

\author{Ronen Weiss}
\affiliation{The Racah Institute of Physics, The Hebrew University, 
             Jerusalem, Israel}
\author{Betzalel Bazak}
\affiliation{Institut de Physique Nucl\'eaire, CNRS/IN2P3, 
Universit\'e Paris-Sud, F-91406 Orsay, France}
\affiliation{The Racah Institute of Physics, The Hebrew University, 
             Jerusalem, Israel}
\author{Nir Barnea}
\email{nir@phys.huji.ac.il}
\affiliation{The Racah Institute of Physics, The Hebrew University, 
             Jerusalem, Israel}

\date{\today}

\begin{abstract} 
Using the zero-range model, it was demonstrated recently
that Levinger's quasi-deuteron model
can be utilized to extract the nuclear neutron-proton contact. 
Going beyond the zero-range approximation and considering the full 
nuclear contact formalism,
we rederive here the quasi-deuteron model for the nuclear 
photoabsorption cross-section and 
utilize it to establish relations and 
constrains for the general contact matrix.
We also define and demonstrate the importance of the
diagonalized nuclear contacts, which can be also 
relevant to further applications of the nuclear contacts.
\end{abstract}

% 67.85.-d - ultracold gases
 %05.30.Fk - Fermion systems and electron gas
 %25.20.-x - Photonuclear reactions
\pacs{67.85.-d, 05.30.Fk, 25.20.-x}

\maketitle
%=============================================================================
% Introduction
%=============================================================================
{\it Introduction --}
For photons in the energy range above the giant resonance,
the nuclear photoabsorption cross-section  
is dominated by two competing mechanisms: the one-body
photomesonic (PM) effect associated mainly with the $M1$ transition,
and the two-body quasi-deuteron (QD) process associated with
the $E1$ transition \cite{TavTer92}.   
The QD process, proposed by Levinger more than 60 years ago 
\cite{Lev51}, is the leading photoabsorption mechanism 
at energies below the pion threshold, and it has a sizable contribution
to the total cross-section up to about 600 MeV 
\cite{Arenhovel91}. Above the pion threshold, the PM effect takes over 
and isobaric excitations combined with meson production become
the dominant features in the cross-section.

Here we focus on the QD process, and more specifically on the relation
between the QD process and the short range correlations
in the nuclear wave function.

In the QD picture, the photonuclear reaction mechanism goes 
through an absorption of the photon by a correlated proton-neutron ($pn$)
pair being close to each other, followed by an emission  
of the two particles flying back to back.
The resulting photonuclear cross-section of a nucleus 
composed of $Z$ protons and $N$ neutrons, $A=N+Z$, is therefore 
expected to be proportional to the deuteron cross-section $\sigma_d$,
\be \label{Levinger}
  \sigma_A(\omega)=L\frac{NZ}{A}\sigma_d(\omega) \;.
\ee
The proportionality constant $L\approx 6$ is known as the
Levinger constant and $\omega$ is the frequency of the photon.
Following Levinger's seminal work, the importance of short range correlations
in nuclear reactions such as
photonuclear reactions \cite{IAEA2000}, hard electron \cite{LO90,FS88,WirSchPie14}
and proton scattering, was realized.
Moreover, the QD process got a remarkable experimental verification
in proton and electron scattering on carbon \cite{Piasetzky06,Subedi08},
and other nuclei \cite{ArrHigRos12,Fom12,Korover14,Hen14},
where high momentum, correlated, $pn$ pairs flying back to back where measured.

When the distance between two particles $r$, is much smaller 
than the average interparticle distance $d_{av}$, the pair's wave
function assumes a specific form that depends  
only on the interaction \cite{Amado72}.
This form assumes an universal $1/r$ shape when the interaction range diminishes 
$R\longrightarrow 0$. Consequently the 
high-momentum tail of the momentum distribution 
fulfills the relation  $\lim_{k\rightarrow\infty}k^4 n(k)= C$. 
Considering a system of two-component fermions interacting 
via such short range  
interaction, Tan \cite{Tan08,Bra12} has established a series of relations 
between the amplitude of the high-momentum tail of the momentum distribution 
$C$, which he coined the ``contact'', and different properties of the system, 
such as the energy, pair correlations and pressure.  
These relations, commonly known as the ``Tan relations'', were 
further extended to other properties and systems by different groups,
see for example \cite{Bra12} and references therein. 
Following the theoretical discovery of the Tan relations,
they were verified in ultracold atomic systems, both in
$^{40}$K \cite{SteGaeDra10,SagDraPau12} and in $^6$Li 
\cite{ParStrKam05,WerTarCas09,KuhHuLiu10} systems.
Moreover, the measured value of the contact, as a function 
of the system's density, was found to be in accordance with the 
theoretical predictions of \cite{WerTarCas09}.

In essence, Tan's contact is a measure of the probability of finding a 
particle pair close to each other. It is therefore not surprising
that the Levinger constant, that counts the number of quasi-deuterons
in a nucleus, is closely related to the contact.
This relation was exposed in \cite{WeiBazBar14} using
the zero-range approximation, and was utilized to evaluate the neutron-proton
contact from the experimental value of the Levinger constant. 

With all its beauty and simplicity, the zero range model
cannot be considered as an accurate description of the nuclear force and
the nuclear wave-function. Within effective field theory (EFT) 
the zero range model is equivalent to the leading order in a pionless theory 
(see e.g. \cite{KSW98,BvK98}), which is known to have a limited range of 
applicability.
Furthermore, realistic nuclear potentials such as $\chi$EFT \cite{Machleidt11,Epelbaum09} 
or AV18 \cite{AV18} that include
pion exchange forces acquire a natural range associated with the pion mass. 
This range is smaller, but not much smaller, than the average nuclear 
interparticle distance.

In view of these limitations of the zero range model, a more general derivation
of the relation between the neutron-proton contact and the Levinger constant is
called for, which is the aim of the current paper. 
To this end we follow the formalism presented in 
\cite{WeiBazBar15a} and utilize it to rederive the QD model.

%=============================================================================
% The Contact in Nuclear Physics
%=============================================================================
{\it The Contact in Nuclear Systems --}
When nucleons $i$ and $j$ come close to each other,
the nuclear wave function $\Psi$ takes on the asymptotic form
\cite{Amado72}
\be\label{full_asymp}
  \Psi\xrightarrow[r_{ij}\rightarrow 0]{}
     \sum_\alpha\varphi_{ij}^\alpha\left(\bs{r}_{ij})
     A_{ij}^\alpha(\bs{R}_{ij},\{\bs{r}_k\}_{k\not=i,j}\right).
\ee
where $\bs{r}_k$ are the single particle coordinates,
$\bs{r}_{ij}=\bs{r}_i-\bs{r}_j$ is the pair's relative distance and
$\bs{R}_{ij}=(\bs{r}_i+\bs{r}_j)/2$ is the center of mass (CM) vector.
The functions $\varphi_{ij}^\alpha$ are called
the asymptotic pair wave functions. They are universal across the nuclear chart
(i.e.
they are independent of the specific nuclear system), and due to symmetry
they only depend on the pair type, i.e.
whether the pair $ij$ is a proton-proton ($pp$) pair, a
neutron-neutron ($nn$) pair or a neutron-proton ($np$) pair.
The sum over $\alpha$ denotes a sum over
the four quantum numbers of the pair $\left(s_2,\ell_2,j_2,m_2\right)$,
which are the pair's total spin $s_2$, its spatial angular momentum
$l_2$ with respect to $\bs{r}_{ij}$, and the total angular momentum and its projection
$j_2$ and $m_2$.
\be
   \varphi_{ij}^{\alpha}\equiv\varphi_{ij}^{(\ell_2s_2)j_2m_2}=
    [\varphi_{ij}^{\{s_2,j_2\}\ell_2}\otimes\chi_{s_2}]^{j_2m_2}
\;,
\ee 
where $\chi_{s_2\mu_s}$ is the two-body spin function, and 
$\varphi_{ij}^{\{s_2,j_2\}\ell_2\mu_\ell}(\bs{r}_{ij})
    ={\phi}_{ij}^{\{\ell_2,s_2,j_2\}} (r_{ij}) Y_{\ell_2 \mu_\ell}(\hat{r}_{ij})$.
Assuming that the nucleus has total angular momentum
$J$ and projection $M$, the matrices of the two-body nuclear contacts
are defined as \cite{WeiBazBar15a}
\be\label{JM_contacts_def}
  C_{ij}^{\alpha \beta}(JM)=16{\pi}^2N_{ij}\bra A_{ij}^\alpha | A_{ij}^\beta \ket.
\ee
Here, $ij$ stands for one of the pairs: $pp$, $nn$ or $np$, 
$N_{ij}$ is the number of $ij$ pairs, and $\alpha$, $\beta$ are the matrix
 indices. In many cases we don't know the nuclear magnetic quantum number.
It is therefore convenient to introduce the 
averaged nuclear contacts, defined as
\be\label{ave_contacts_def}
  C_{ij}^{\alpha \beta}=\frac{1}{2J+1}\sum_M C_{ij}^{\alpha\beta}(JM).
\ee
The averaged contacts $C_{ij}^{\alpha\beta}$
do not depend on $m_\alpha$
or $m_\beta$, but only on $(s_{\alpha},\ell_{\alpha},j_{\alpha})$, and 
$(s_{\beta},\ell_{\beta},j_{\beta})$.
The contacts $C_{ij}^{\alpha \beta}$ are the generalized nuclear analogs of
Tan's contact \cite{Tan08}.

{The factorized asymptotic form given in Eq. (\ref{full_asymp}) should be
satisfied in the limit $r_{ij}\rightarrow 0$ but its exact range
of validity is not fully understood in the available studies.
Such a factorization is the basis for many of the Tan relations
\cite{Tan08,Bra12}.
The relevant length scales in nuclear systems are the
average distance between two nucleons, and the scattering
lengths. It is reasonable to assume that $r_{ij}$ should 
be smaller than these length scales for the factorization
to be valid. Furthermore, using the variational
Monte Carlo (VMC) results of Wiringa {\it et. al.}
\cite{WirSchPie14},
one can estimate that this asymptotic form is
valid for $r_{ij}$ smaller than about $1$ to $2$ fm.
These VMC results were calculated using the Argonne v18
two-nucleon and Urbana X three-nucleon potentials
for $A \leq 12$ nuclei. As the inner parts of these nuclei already possess
the nuclear saturation density we expect that the above estimate will hold for
heavier nuclei.}

{
The above asymptotic factorization does not
take into acount three-body correlations. 
In the zero-range model, the asymptotic form
for the case where three particles approach each other
is given in Ref. \cite{WerCas12}. The contribution of three-body
correlations is expected to be less significant than the
contribution of two-body correlations \cite{BraKanPla11}. Thus,
we will not consider here three-body correlations, and it is left
for future studies. 
}

As will be clear later, it will be useful to work
in a basis for which the contact matrices are diagonal.
$C_{ij}(JM)$ is an Hermitian matrix and therefore can be diagonalized.
So, there exists an unitary matrix $U$
(generally, $U$ depends on the type of the pair $ij$,
on the nucleus and its quantum numbers $J$ and $M$)
and a diagonal matrix $D_{ij}(JM)$ such that
\be
   D_{ij}(JM)=UC_{ij}(JM)U^{-1}
\ee
We can also define
\be
\tilde{\varphi}_{ij}^\alpha=\sum_\beta U_{\alpha\beta}\varphi_{ij}^\beta
\ee
and
\be
\tilde{A}_{ij}^\alpha=\sum_\beta (U^{-1})_{\beta\alpha}A_{ij}^\beta=
\sum_\beta U^*_{\alpha\beta} A_{ij}^\beta.
\ee
It is now simple to prove that
\be
\sum_\alpha \varphi_{ij}^\alpha A_{ij}^\alpha = 
\sum_\alpha \tilde{\varphi}_{ij}^\alpha \tilde{A}_{ij}^\alpha
\ee
and
\be\label{diag_contact}
   16\pi^2N_{ij} \langle \tilde{A}_{ij}^\alpha | \tilde{A}_{ij}^\beta \rangle 
       = \delta_{\alpha\beta} D_{ij}^{\alpha\alpha} (JM).
\ee
This way we have defined here a new
basis for which the contact matrices are diagonal.
In some sense it is not the natural basis to work with,
because the $\tilde{\varphi}_{ij}^{\alpha}$ are not
universal as they depend on the specific nucleus (because
of $U$). Nevertheless, this basis will be very useful
to our purpose of rederiving the QD model.

The averaged diagonal contacts can also be defined
\be
  D_{ij}^{\alpha \alpha}=\frac{1}{2J+1}\sum_M D_{ij}^{\alpha\alpha}(JM).
\ee

%=============================================================================
% The quasi-deuteron model
%=============================================================================
{\it The Quasi-Deuteron model -- }
%In the following we will relate the contacts
%to the quasi-deuteron model. 
In the leading $E1$ dipole  approximation, the total photo absorption cross section of 
a nucleus is given by 
\begin{equation}\label{cross_section_A}
  \sigma_A(\omega)=4\pi^{2}\alpha\hbar\omega R(\omega)\,,
\end{equation}
\noindent where $\alpha$ is the fine structure constant, and
\begin{equation} \label{1}
  R(\omega )=\bar{\sum_i} \sum_{f}\left| \bra \Psi _{f}\right| \bs{\epsilon}\cdot
   \hat{\bs D} \left| 
\Psi _{0}\ket \right| ^{2}\delta (E_{f}-E_{0}-\hbar\omega) 
\end{equation}
is the response function. $\hat{\bs D}$ is the unretarded dipole operator
$\hat{\bs D}=\sum_{i=1}^{A}\frac{1+\tau^{3}_{i}}{2}\bs r_{i}$, 
and ${\bs \epsilon}$ is the photon's polarization vector.
The initial (ground) state and the final state wave functions are denoted by 
$\left| \Psi_{0/f} \right\ket$ and the energies by $E_{0/f}$, respectively. 
The operator $\tau^{3}_{i}$ is the third component of the $i$-th nucleon 
isospin operator. 
The sum $\sum_f$ in the response function is a sum over
the final states that becomes an integration in the limit of infinite volume.
The response function also contains an average over
the initial states which amounts to an average over the 
magnetic projection of the ground state, $\bar\sum_i=1/(2J_0+1)\sum_{M_0}$.
We note that the different final states must be orthogonal eigenstates
of the nuclear Hamiltonian. For that reason it will be important to work
with the diagonalized contact matrices.

%$\Psi_0$ obeys the asymptotic form of Eq. (\ref{full_asymp}).

For {inverse photon wave number $q^{-1}$} somewhat shorter than the average
interparticle distance ($q d_{av} > 1$),
the photon is absorbed by a QD pair.
Consider a reaction mechanism 
where the photon is absorbed by a proton $p$ that is emitted with large momentum
$\mathbf{k}_{p}$. For high photon energies this process is fast
enough so we can use the Born approximation,
which means that any interaction between the emitted
proton and the rest of the nucleus can be neglected.
Hence, due to momentum conservation, another particle must be emitted. 
As pointed out by Levinger \cite{Lev51}, because of
the $E1$ nature of the process, this other particle
must be a neutron $n$,
since proton-proton pair posses no dipole moment. 
The emitted neutron's momentum $\mathbf{k}_{n}$ is such that
$\mathbf{k}_{n}\approx-\mathbf{k}_{p}\equiv\mathbf{k}$, thus
the relative momentum of the emitted pair is 
$\frac{\mathbf{k}_{n}-\mathbf{k}_{p}}{2}=\frac{2\mathbf{k}_{n}}{2}=\mathbf{k}$.
This point can 
be further reinforced, under the 
assumption that the center of mass obeys $\sum_{i=1}^A \bs{r}_i=0$,
through the algebraic relation
\be
  \hat{\bs D}=\frac{1}{A}
      \sum_{i,j=1}^{A}\frac{\tau^{3}_{i}-\tau^{3}_{j}}{4}(\bs r_{i}-\bs r_{j}) \;,
\ee
explicitly writing the dipole operator as a two-body operator
that vanishes for all but $np$ pairs. Nevertheless, it will be
more convenient to work with
the one-body representation of the dipole operator in the
following derivations.
The relative motion of the emitted pair contains most of the photon energy, whereas
the photon's momentum is translated into the CM motion. The energy fraction 
associated with the CM coordinate $\bs{R}_{pn}$ is $\hbar\omega/4 Mc^2$,
{where $M$ is the mass of the nucleons},
which is only few percents for the photon energies under consideration. 
We can therefore safely neglect the pair's recoil.

Assuming that the residual $A-2$ particles wave function is frozen throughout 
this process, we can write the final state wave functions for the channel $\alpha$,
normalized in a box of volume $\Omega$, 
in the following way
\be \label{eq:final state}
\Psi_{f}^{\alpha s \mu_s}={\cal N}_\alpha
      \hat{\cal A}\left\{\frac{1}{\sqrt{\Omega}}e^{-i\mathbf{k}\cdot\mathbf{r}_{pn}}
       \chi_{s,\mu_s}
       \tilde{A}^\alpha_{pn}(\bs{R}_{pn},\{\bs{r}_{j}\}_{j\neq p,n})\right\}\;.
\ee
Here ${\cal N}_\alpha$ is a normalization factor, 
and
$ \hat{\cal A}=\big(1-\sum_{p'\neq p}(p,p')\big)
             \big(1-\sum_{n'\neq n}(n,n')\big) $
is the proton-neutron antisymmetrization operator with $(i,j)$ being the 
transposition operator. 
The sums over $p', n'$ extends over all protons and neutrons in the system but 
$p, n$.
As $A^\alpha_{pn}(\bs{R}_{pn},\{\bs{r}_{j}\}_{j\neq p,n})$ is antisymmetric under 
permutation of all identical particles but the pair $pn$, $\Psi_{f}^{\alpha s \mu_s}$ is
antisymmetric under proton permutations and under neutron permutations.
Note that the different final state functions are indeed 
orthogonal as required. Here the importance of the ``diagonal basis",
the $\tilde{A}$ functions, becomes clear.
{If we were to use the original functions $A$
 in the definition of the final states,
then they would not have been orthogonal
and the whole derivation would have
been wrong.}
The normalization factor is given by
\be\label{norm-alpha}
  {\cal N}_\alpha=\frac{1}{\sqrt{NZ}}
       \frac{1}{\sqrt{\bra \tilde{A}^\alpha_{pn}|\tilde{A}^\alpha_{pn}\ket}}
  =\frac{4\pi}{\sqrt{D_{pn}^{\alpha\alpha}(J_0 M_0)}}.
\ee

Considering now the transition matrix element we see that
\begin{align}
\bra & \Psi_{f}^{\alpha s \mu_s}|\bs{\epsilon}\cdot\hat{\bs D}|\Psi_{0}\ket 
    =  N Z {\cal N}_\alpha 
    \int \prod_{k}d\bs{r}_{k} \frac{1}{\sqrt{\Omega}} \cr
    & \times
    e^{i\mathbf{k}\cdot\mathbf{r}_{pn}}
    \chi_{s,\mu_s}^\dagger
    \tilde{A}_{pn}^{\alpha\dagger}\left(\mathbf{R}_{pn},\{\mathbf{r}_{j}\}_{j\neq pn}\right)
    \left(\bs{\epsilon}\cdot\hat{\bs D}\right)\Psi_{0}
\end{align}
where we have used the fact that $\hat{\cal A}\Psi_{0}=NZ\Psi_{0}$.
Due to the orthogonality of the initial and final states, the transition 
matrix element vanishes unless the photon acts on the outgoing $pn$ pair.
The above integral contains the oscillatory function
$e^{i\mathbf{k}\cdot\mathbf{r}_{pn}}$,
so since the momentum $\mathbf{k}$ is large
only the asymptotic part $r_{pn}\rightarrow 0$
will contribute to the integral. Therefore
the integration over $r_{pn}$ hereafter can be limited to a small neighborhood 
of the origin $\Omega_0$, {where the asymptotic form
of the wave function (\ref{full_asymp}) is valid (for more details
see Ref. \cite{WeiBazBar14} and especially its supplemental materials).} 
Hence,
\begin{align}\label{D_me2}
\bra \Psi_{f}^{\alpha s \mu_s}|\bs{\epsilon}\cdot\hat{\bs D}|\Psi_{0}\ket 
    &
    = N Z {\cal N}_\alpha \sum_{\beta} \bra \tilde{A}^\alpha_{pn} | \tilde{A}^{\beta}_{pn} \ket 
     \cr & \times 
    \bra \bs{k} s \mu_s | \bs{\epsilon}\cdot\hat{\bs D}_{pn} | \tilde\beta \ket \;,
\end{align}
where
\be\label{me_2_tilde}
    \bra \bs{k} s \mu_s | \bs{\epsilon}\cdot\hat{\bs D}_{pn} | \tilde\beta \ket = 
    \int_{\Omega_0} d\bs{r}\frac{1}{\sqrt{\Omega}}e^{i\mathbf{k}\cdot\mathbf{r}}    
     \chi_{s,\mu_s}^\dagger 
    \bs{\epsilon}\cdot\hat{\bs D}_{pn}
    \tilde{\varphi}_{pn}^\beta(\bs{r}) \;,
\ee
and $\hat{\bs{D}}_{pn}=\frac{\mathbf{r}_{pn}}{2}$.
Working in the "diagonal basis"
only $\beta=\alpha$ contributes to the sum in (\ref{D_me2}). 
Substituting now the normalization factor (\ref{norm-alpha})
and utilizing the relation (\ref{diag_contact}), 
\begin{align}\label{me_A_tilde}
   \bra\Psi_{f}^{\alpha s \mu_s}|\bs{\epsilon}\cdot\hat{\bs D}|\Psi_{0}\ket &
    = \frac{\sqrt{D_{pn}^{\alpha \alpha}(J_0 M_0)}}{4\pi}
    \bra \bs{k} s \mu_s | \bs{\epsilon}\cdot\hat{\bs D}_{pn} | \tilde\alpha \ket \;.
\end{align}
If we now express $\tilde{\varphi}_{pn}^\alpha(\bs{r})$ through the 
asymptotic pair wave functions we get
\begin{align}\label{me_A}
  \bra\Psi_{f}^{\alpha s \mu_s}|\bs{\epsilon}\cdot\hat{\bs D} &|\Psi_{0}\ket
    = \frac{\sqrt{D_{pn}^{\alpha \alpha}(J_0 M_0)}}{4\pi}
\cr & \times 
    \sum_\beta U_{\alpha\beta}
    \bra \bs{k} s \mu_s | \bs{\epsilon}\cdot\hat{\bs D}_{pn} | \beta \ket 
\end{align}
where 
\be\label{me_2}
    \bra \bs{k} s \mu_s | \bs{\epsilon}\cdot\hat{\bs D}_{pn} | \alpha \ket 
      = 
    \int_{\Omega_0} d\bs{r}\frac{1}{\sqrt{\Omega}}e^{i\mathbf{k}\cdot\mathbf{r}}    
     \chi_{s,\mu_s}^\dagger 
    \bs{\epsilon}\cdot\hat{\bs D}_{pn}
    {\varphi}_{pn}^\alpha(\bs{r}) \;.
\ee

We can now take the square absolute value of this matrix element,
average over the initial states and sum over final states.
The sum over the final states is a sum over $\alpha$, $s$ and $\mu_s$, 
and integration over $\bs{\hat{k}}$. Starting with the sum over $\alpha$, $s$ and $\mu_s$
we get
\begin{align}\label{squared_me}
   \sum_{\alpha s \mu_s}&\left| \bra \Psi_{f}^{\alpha s \mu_s}| \bs{\epsilon}\cdot
       \hat{\bs D} | \Psi_{0}\ket \right|^{2}
    = \cr
    = & \sum_{\alpha}\frac{{D_{pn}^{\alpha \alpha}(J_0 M_0)}}{16\pi^2} 
    \sum_{s, \mu_s}
    \left| \sum_\beta U_{\alpha\beta}
    \bra \bs{k} s \mu_s | \bs{\epsilon}\cdot\hat{\bs D}_{pn} | \beta \ket \right|^2
   \cr
    = & \sum_{\beta,\beta'}
   \frac{ \sum_{\alpha} U_{\alpha\beta}^*D_{pn}^{\alpha \alpha}(J_0 M_0)U_{\alpha\beta'}}{16\pi^2}
    R_{\beta\beta'}(\bs{k}) \cr
    = & \sum_{\beta,\beta'} \frac{C_{pn}^{\beta\beta'}(J_0 M_0)}{16\pi^2}
         R_{\beta\beta'}(\bs{k})
\end{align}
where
\be
    R_{\beta\beta'}(\bs{k})= 
     \sum_{s,\mu_s}
     \bra \bs{k} s \mu_s | \bs{\epsilon}\cdot\hat{\bs D}_{pn} | \beta \ket^*
     \bra \bs{k} s \mu_s | \bs{\epsilon}\cdot\hat{\bs D}_{pn} | \beta' \ket
\ee
Integrating now over the momentum $\bs{\hat{k}}$ and averaging over the initial states
we get
\be\label{response_fun}
   R(\omega) = \sum_{\beta,\beta'} \frac{C_{pn}^{\beta\beta'}}{16\pi^2}
        R_{\beta\beta'}(\omega) \;,
\ee
where
\be
  R_{\beta\beta'}(\omega)=\int \frac{d\hat{\bs{k}}}{(2\pi)^3}R_{\beta\beta'}(\bs{k})\;.
\ee
Deriving (\ref{response_fun}), we have utilized the 
sum over $M_0$ and replaced the contact matrix
$C_{pn}(J_0M_0)$, with the averaged contacts $C_{pn}$.
This step could not have been done before, as in general
the matrix $U$ depends on the specific nucleus and its
quantum numbers $J_0$ and $M_0$. 

The response function (\ref{response_fun}) is 
a general result valid for all nuclei. It is composed 
of a particular part,
that depends on the specific nucleus through the values
of the contacts, and an universal part $R_{\beta\beta'}(\omega)$
that doesn't change along the nuclear chart. 
{It is universal since it is written using the
original and physical $\varphi$ functions and 
does not include the $\tilde{\varphi}$ functions nor the matrix
$U$. As explained before, it was necessary to use
the diagonal basis in the derivation, but it is more useful
to write the final result using the physical basis}.
We notice that only $\beta$ and $\beta'$
with $s_\beta=s_{\beta'}$ can contribute to the response
(even though generally the contacts
are not diagonal in $s$) because 
$\hat{\bs D}_{pn}$ is a spin scalar, and therefore if 
 $s_\beta\neq s_{\beta'}$ then all terms in the sum over $s,\mu_s$
must vanish.

{Eq. \eqref{response_fun} should hold when 
$\omega \rightarrow \infty$ and its exact range of validity is
directly connected to the validity range of the
asymptotic form \eqref{full_asymp}. If Eq. \eqref{full_asymp}
holds for $r$ smaller then some distance $d_{a}$ then we would expect that Eq.
\eqref{response_fun} would hold for $q d_{a} > 1$,
where $q$ is the photon wave number. As mentioned before,
$d_{a}\approx 1-2$ fm according to the VMC data
of Wiringa {\it et. al.} \cite{WirSchPie14}.
Thus we expect Eq. \eqref{response_fun} to hold for
\be
   \hbar \omega =\hbar q c > \frac{\hbar c}{d_{a}}\approx 100 - 200 \mathrm{MeV}.
\ee
We note that the E1 transition considered here
is the leading effect
up to about $140$ MeV and the extraction of the 
Levinger constant was usually done from experiments with
photon energies between $40$ MeV and $140$ MeV \cite{TavTer92}.
In order to use experimental data with higher energies,
a separation of the E1 transition from the total photoabsorption
cross section is needed. 
}

%=============================================================================
{\it The relation to the QD model --}
The result (\ref{response_fun}) is also valid for the deuteron.
The deuteron is a bound proton-neutron pair 
with angular momentum $J=1, M=0,\pm 1$, positive parity,
and total spin $S=1$.
Since it is only a two body system, the quantum numbers of
the full state determine that many of the deuteron contacts are zero.
$A_{pn}^\alpha$ can be different from zero
only for $\alpha=(l=0,s=1,j=1,m=M)$ and $\alpha=(l=2,s=1,j=1,m=M)$.
So, given a projection $M$, we have only one $2 \times 2$
block of non-zero contacts. Each $M$ defines a different block,
however these blocks must be identical.
The deuteron averaged contact is therefore composed of three
identical $2 \times 2$ blocks.
%As mentioned in our last paper (Generalized nuclear
%contacts and the nucleon's momentum distributions)
%the averaged contacts $C_{ij}^{\alpha\beta}$
%do not depend on $m_\alpha$
%or $m_\beta$, and here the indices of all these three blocks
%differ only by $m=\pm1,0$. Therefore, these three
%blocks are identical. As we explained before,
%these blocks are just the blocks that exist in
%each of the non-averaged matrices of contacts
%(up to a factor of $1/3$). It means that each
%non-averaged matrix (for each $M$) is built from one
%$2 \times 2$ block and it is the same block for all
%three matrices (but it is located in different indices
%for each matrix). This discussion is of course
%relevant only for the deuteron.
%$Concluding, for the deuteron there are generally
%12 non-zero contacts, that can be separated
%into 4 sets of 3 contacts. Each set contains three
%contacts $C_{pn}^{\alpha\beta}$ with the same value,
%all the three are different only by the value of $m_\alpha$
%and $m_\beta$ $(m_\alpha=m_\beta=\pm1,0)$. 
The deuteron contact $C_{pn}^{\alpha\beta}$ is defined by the
values of $\ell_\alpha$ and $\ell_\beta$ (being zero or two).
%The values of the contacts with $(\ell_\alpha=0,\ell_\beta=2)$
%and with $(\ell_\alpha=2,\ell_\beta=0)$ are
%the complex conjugate of each other.

Comparing the Levinger model \eqref{Levinger}
with the response function \eqref{response_fun}, we see that the latter
is complex and involve contact channels that are missing in the deuteron.
{As a result, we must conclude that the QD model cannot be completely accurate,
 since heavy nuclei include two-body channels
that does not exist in the deuteron.
Nevertheless, the QD model does describe nicely the available experimental
data. As a result, we can obtain approximated constrains on the different
nuclear contacts within the same accuracy.}
In the following we'll explore these implications, and analyze under 
what conditions the nuclear photoabsorption cross-section 
becomes proportional to the deuteron's.

The $pn$ contact $C_{pn}^{\alpha\beta}$ measures
the probability to find a neutron close to a proton in 
the specific $\alpha\beta$ channel. 
As nuclei behave as an incompressible liquid having almost constant density, 
it seems reasonable
that the way these probabilities scale with the number of
protons, $Z$, and neutrons, $N=A-Z$, does not depend on the specific channel.
It means that if in nucleus $Y$ the probability to find an $np$
pair in channel A is twice the probability to find it in nucleus $X$,
then also in channel B the probability is twice in nucleus $Y$
than in nucleus $X$. If this is the case, then it means that
the $pn$ contacts scale with the number of proton and neutrons
regardless of $\alpha$ or $\beta$. It means that there exist
$\eta_{pn}(N,Z)$, independent of $\alpha$ or $\beta$, such that
\be \label{scaling}
C_{pn}^{\alpha\beta}(^A_Z\!X)=\eta_{pn}(N,Z)C_{pn}^{\alpha\beta}(d),
\ee
where $X$ is a nucleus in its ground state and $d$ is the deuteron.
This should only be an approximate relation because
we don't expect that all the contacts that are exactly zero
for the deuteron will also be exactly zero in heavier nuclei.
%\red{As explained before, we don't expect the QD model to 
%be exact, but this is one way to explain the fact that this
%model is a good approximation.}
This relation also includes the case where there is only one
significant contact for all the nuclei (one significant channel
for a neutron to be close to a proton).
So, if we assume this relation given in Eq. (\ref{scaling}),
we get directly from Eq. \eqref{response_fun} that
\be
\sigma_X(\omega)=\eta_{pn}(N,Z) \sigma_d(\omega),
\ee
where Z and N are the number of protons and neutrons
in the nucleus $X$. We can compare it to
Eq. (\ref{Levinger}) and get
\be
  \eta_{pn}=L\frac{NZ}{A}.
\ee

There might also be a different scenario that yields
a constant ratio between the cross sections. 
As can be seen in Eq. \eqref{response_fun},
each contact is multiplied by some universal function of $\omega$.
If these functions are proportional to each other, or at least that is the
case for the dominant ones, there will be a constant ratio between the 
photoabsorption cross-sections. 
%Assuming that there are for example two contacts (or more) that are
%multiplied by the same universal function of $\bs{k}$ and that they are the most
%significant contacts (the rest are negligible). Also we assume that at least
%one of this contacts is the most significant also in the deuteron
%(and that all the significant contacts in the deuteron also
%come with a universal function of $\omega$).
In this case we will get
a constant ratio regardless of the scaling of the contacts
with Z and N because the $\omega$-dependence is canceled.
This is actually the situation in the zero-range model
assumed in \cite{WeiBazBar14}.
From the zero-range model it follows that the only non-zero
deuteron contact is the s-wave spin-triplet contact.
For heavier nuclei, both s-wave spin-singlet
and spin-triplet might be significant, but are assumed to
come with the same asymptotic pair wave function $\propto 1/r$.
In this scenario we get that
\be
  \frac{\sum'_{\alpha,\beta}C_{pn}^{\alpha,\beta}(^A_Z\!X)}
       {\sum'_{\alpha',\beta'}C_{pn}^{\alpha',\beta'}(d)}=L\frac{NZ}{A}
\ee
where the notation $\sum'$ indicates that the sum is restricted to the dominant
channels. In general, we expect these channels to be purely
or partially s-wave channels.

{When further details regarding
the values of the different nuclear contacts
become available, it will be possible to use Eq. 
\eqref{response_fun} to deduce the 
corrections to the QD model}.

%==============================================================================
{\it Conclusions --}
Summing up,
we have rederived here the QD model using the full
nuclear contact formalism. Assuming that Levinger's model is accurate we
have obtained few constrains on the $np$ contact matrix,
and on the scaling of the $np$ contacts along the nuclear chart.
We have also defined here the diagonalized nuclear
contacts and emphasized their importance in the
derivation presented in this manuscript. The diagonalized
contacts might turn out to be an important
tool in future derivations of the nuclear contact relations.

%==============================================================================
\begin{acknowledgments}
{This work was supported by the Pazy Foundation.}
\end{acknowledgments}
%==============================================================================

%==============================================================================
\end{document}